%% file: main.tex
\documentclass[pra,showpacs,showkeys,twocolumn]{revtex4-1}

\usepackage{graphicx}
\usepackage{latexsym}
\usepackage{amsmath}
\usepackage{amssymb}
\usepackage{amsfonts}
\usepackage{color}
\usepackage[dvipsnames]{xcolor}
\usepackage{textcomp}
\usepackage[normalem]{ulem}

\newcommand{\xx}{\mathbf{x}}
\newcommand{\kk}{\mathbf{k}}

\begin{document}
\title{Noise-induced transition from superfluid to vortex state in
two-dimensional nonequilibrium polariton condensates --
semi-analytical treatment}
\author{Vladimir N. Gladilin and Michiel Wouters}

\affiliation{TQC, Universiteit Antwerpen, Universiteitsplein 1,
B-2610 Antwerpen, Belgium}

\begin{abstract}
    We develop a {semi-}analytical description for the
Berezinskii-Kosterlitz-Thouless (BKT) like phase transition in
nonequilibrium Bose-Einstein condensates. { Our theoretical analysis
is based on a noisy generalized Gross-Pitaevskii equation. Above a
critical strength of the noise, spontaneous vortex-antivortex pairs
are generated. We provide a semi-analytical determination of the
transition point} based on a linearized Bogoliubov analysis, to
which some nonlinear corrections are added. We present two different
approaches that are in agreement with our numerical calculations in
a wide range of system parameters. {We find that for small losses
and not too small energy relaxation, the critical point approaches
that of the equilibrium BKT transition. Furthermore, we find that
losses tend to stabilize the ordered phase: keeping the other
parameters constant and increasing the losses leads to a higher
critical noise strength for the spontaneous generation of
vortex-antivortex pairs. } Our theoretical analysis is relevant for
experiments on microcavity polaritons.
\end{abstract}
\date{\today}

\maketitle

\section{Introduction}

The interest in nonequilibrium phase transitions of quantum many
body systems has witnessed a rapid growth over the last decade
thanks to the developments in Bose-Einstein condensation in optical
systems (microcavity polaritons and photons in dye filled cavities)
\cite{bloch2022non}, circuit QED \cite{carusotto2020photonic} and
ultracold atomic gases \cite{labouvie_bistability_2016}. One of the
most elementary phase transitions in these systems is the onset of
Bose-Einstein condensation, defined as the emergence of spontaneous
long range phase coherence. Where at thermal equilibrium, long range
phase coherence appears when the temperature is lowered below a
density-dependent critical temperature, in nonequilibrium systems,
the phase coherence is determined by the interplay between the
hamiltonian and dissipative parts of the dynamics or even between
competing dissipative mechanisms
\cite{vanregemortel2021entanglement,diehl2008quantum}.

Since quantum fluids of light are only available in one or two
dimensions, true long range order is actually absent. In
one-dimensional bose gases, both at thermal equilibrium and out of
equilibrium, the spatial decay of the first order coherence function
is always exponential \cite{wouters2006absence,chiocchetta2013non}.
In two dimensions and at equilibrium there is the celebrated
Berezinskii-Kosterlitz-Thouless phase transition
\cite{berezinskii1971destruction,kosterlitz1973ordering} that
separates the normal and the superfluid state, with exponential and
algebraic decay of the spatial coherence respectively. {In
equilibrium, the phase dynamics is in the XY universality class and
the corresponding universal jump in the superfluid stiffness has
been experimentally observed in $^4$He~\cite{bishop1978study}. More
recently, the flexibility of the platform of ultracold atoms allowed
a direct observation of the spontaneous formation of
vortex-antivortex pairs above the BKT transition
\cite{hadzibabic2006berezinskii}. The ultracold atomic gases are in
the weakly interacting regime, for which the transition temperature
was computed by Prokof'ev and Svistunov by a clever combination of
the linear Bogoliubov approximation and numerical Monte Carlo
simulations~\cite{prokof2001critical}}.

{For photonic systems} out of equilibrium, the phase dynamics is
actually in the Kardar-Parisi-Zhang universality class where a
nonlinear term in the phase evolution is essential
\cite{wachtel2016electrodynamic,ji2015temporal}. For one-dimensional
polariton systems, the spatial decay of the correlations remains
qualitatively unaffected by the nonlinearity in the phase dynamics
\cite{gladilin2014spatial}, but a specific spatiotemporal scaling
emerges, that was recently observed experimentally
\cite{fontaine2022kardar}.

In two dimensions, the KPZ phase dynamics was predicted to make long
range phase coherence impossible in isotropic systems
\cite{altman2015two,wachtel2016electrodynamic}. Numerical studies on
the other hand have shown a transition toward a state with algebraic
decay of the coherence \cite{dagvadorj2015nonequilibrium} and an
associated disappearance of vortex-antivortex pairs
\cite{dagvadorj2015nonequilibrium,caputo2018topological,gladilin2019noise-induced,dagvadorj2022unconventional}
without the formation of topological defects even when the
spatiotemporal correlations feature KPZ scaling
\cite{mei2021spatiotemporal,deligiannis2022kardar}. Since
computational resources limit the system sizes for numerical
studies, the discrepancy between the renormalisation group studies
could be due to finite size effects, but at present it does not seem
that the issue is fully settled. Even when the numerically observed
BKT transition is due to a limited system size, experimentally
available systems necessarily also work with relatively small sizes,
so that there is a clear interest in the nonequilibrium BKT
transition. {Compared to the equilibrium case, the current
understanding of the dependence of the BKT critical point on the
system parameters is much less mature. The reason herefore is
twofold. First, out of equilibrium the standard Boltzmann-Gibbs
ensemble can no longer be used and the steady state has to be
characterized by a more involved simulation of the system dynamics.
Second, the nonequilibrium dynamics is governed by more parameters:
in addition to the system Hamiltonian and environment temperature,
also the details of the coupling to the environment come into play
in the non-equilibrium situation.}

{In our previous work on photon
condensation~\cite{gladilin2021vortex}, we have pinpointed the
nonequilibrium BKT critical point with numerical simulations and
developed a semi-analytical approach in order to get a better
understanding of the location of the critical point. In our
numerical simulations, the transition was approached from the
ordered side with no vortices present in the initial state. Above a
critical value of the noise strength in the stochastic classical
field description of the dynamics, vortex-antivortex pairs
spontaneously appear, signalling the BKT like transition to the
disordered state.} {Our work involved both} numerical simulations
and  analytical approximations that capture the dependences of the
transition point on all the system parameters. The analytical
approximation for photon condensates was based on the Bogoliubov
approximation, combined with an infrared cutoff set by the inverse
vortex core size \cite{gladilin2020vortices}. In our previous study
on the BKT transition for (interacting) polaritons
\cite{gladilin2019noise-induced}, no such analytical estimate was
given.

In the present article, we wish to fill this gap. Moreover, we
extend our previous results to the regime of vanishing interactions,
so that we can elucidate the effect of both the nonequilibrium
condition and of interactions on the BKT transition point. When the
interactions become small compared to the gain saturation
nonlinearity, the vortex core size can significantly deviate from
the usual healing length defined as $\xi = \hbar/\sqrt{mg\bar n}$,
where $m$ is the mass, $g$ the interaction constant and $\bar n$ the
density of polaritons in the condensate. The vortex core size
appears in our treatment as a good proxy for the inverse of the
infrared cutoff that we have to introduce to avoid the divergence of
a momentum integral. We therefore carried out a systematic analysis
of the vortex size and structure as a function of the strength of
the interactions and of the driving and dissipation.

The structure of this paper is as follows. In Sec. \ref{sec:model},
we introduce our model for polariton condensates and derive the
density and phase flucutations within the linear (Bogoliubov)
approximation. In Sec. \ref{sec:approx}, we construct some
approximate formulae for the BKT critical point with a few fitting
parameters that are able to capture our numerical simulations. We
start with a simple approach that is able to capture the main
dependencies of the critical point on the system parameters and then
present a more refined approach that allows for a very good fitting
of the numerical results. Conclusions are drawn in Sec.
\ref{sec:concl} and the vortex structure is discussed in appendix
\ref{app:vort}.

\section{Model and linearization \label{sec:model}}

We consider nonresonantly excited two-dimensional polariton
condensates. In the case of sufficiently fast relaxation in the
exciton reservoir, this reservoir can be adiabatically eliminated
and the condensate is described by the noisy generalized
Gross-Pitaevskii
equation~\cite{Wouters:PRB2009,Szymanska:PRB2007,sieberer2016keldysh,carusotto:2013}
\begin{eqnarray}
({\rm i}-\kappa)\hbar \frac{\partial \psi}{\partial t} =&&
\left[-\frac{\hbar^2\nabla^2}{2m} +g |\psi|^2 \right. \nonumber
\\
&&\left.+\frac{{\rm i}}{2} \left(\frac{P}{1+|\psi|^2/n_s}-\gamma
\right) \right] \psi+\sqrt{D}  \xi . \label{ggpe}
\end{eqnarray}
Here $m$ is the effective mass and the contact interaction between
polaritons is characterized by the strength $g$. The imaginary term
in the square brackets on the right hand side describes the
saturable pumping (with strength $P$ and saturation density $n_s$)
that compensates for the losses ($\gamma$). We take into account the
energy relaxation $\kappa$ in the
condensate~\cite{wouters2012energy}. The complex stochastic
increments have the correlation function $\langle \xi^*(x,t)
\xi(x',t') \rangle=2 \delta({\bf r}-{\bf r}') \delta(t-t')$.
{Eq.\eqref{ggpe} is a classical stochastic field model that
describes all the fluctuations in the system as classical. This
model is therefore only valid in the weakly interacting regime
$gm/\hbar^2 \ll 1$, where quantum fluctuations are small.}

For $\kappa=0$, the zero momentum steady state of Eq. \eqref{ggpe}
is under homogeneous pumping $\psi_0(\xx,t) = \sqrt{n_0} e^{-i g n_0
t}$, with $n_0= n_s(P/\gamma-1 )$. By expressing the particle
density $|\psi|^2$ in units of $n_0$, dividing time by
$\hbar(1+\kappa^2)/n_0$, length by $\hbar/\sqrt{2mn_0}$, and noise
intensity by $\hbar^3 n_0/(2m)$, Eq. \eqref{ggpe} takes the form:
\begin{align}
\frac{\partial\psi}{\partial t}=& (i+\kappa)\left[\nabla^2
-g|\psi|^2 -\frac{i\gamma}{2n_s}\frac{1-|\psi|^2}{1+\nu |\psi|^2}
\right] \psi \nonumber \\ &+\sqrt{D} \xi, \label{ggpe2}
\end{align}
where $\nu = n_0/n_s$. The steady state density is then in the
absence of noise given by \cite{gladilin2019noise-induced}
\begin{align}
\bar n= \sqrt{\left(\frac{\kappa+c}{2\kappa\nu}\right)^2+
\frac{c}{\kappa\nu}}-\left(\frac{\kappa+c}{2\kappa\nu}\right)
 \label{barn}
\end{align}
with $c\equiv\gamma/(2gn_s)$.

In order to gain some insight in the physics of the fluctuations
induced by the noise in Eq. \eqref{ggpe2}, one can consider in first
approximation the linearized equations for the density and phase
fluctuations around the steady state:
\begin{equation}
    \psi(\xx,t) = \sqrt{\bar n + \delta n(\xx,t)} e^{-i g \bar n t + i \delta \theta(\xx,t)}
\end{equation}
After a spatial Fourier transform, these obey the linearized
equations of motion
\begin{align}
\frac{\partial}{\partial t} \delta \theta_\kk &= -\kappa
\epsilon_\kk \delta \theta_\kk -\frac{\epsilon_\kk}{2 \bar n} \delta
n_\kk -(g-\kappa \tilde \gamma)\delta n_\kk \nonumber \\ &+
\sqrt{\frac{D}{\bar n}} \xi^{(\theta)}_\kk, \label{dph}
\end{align}
\begin{align}
\frac{1}{\bar n}\frac{\partial}{\partial t} \delta n_\kk &= -\kappa
\epsilon_\kk  \frac{\delta n_\kk }{\bar n}+2\epsilon_\kk \delta
\theta_\kk -2(\kappa g +\tilde \gamma)\delta n_\kk \nonumber
\\ &+ 2\sqrt{\frac{D}{\bar n}} \xi^{(n)}_\kk, \label{dn}
\end{align}
where
\begin{align}
\tilde \gamma=\frac{\gamma(1+\nu)}{2n_s(1+\nu\bar n)^2}.
\label{tgam}
\end{align}

Using the Ito formula \cite{jacobs2010stochastic}, one can obtain
from Eqs. (\ref{dph}) and (\ref{dn}) a set of three equations:
\begin{align}
\frac{D}{\bar n\epsilon_\kk}&=2\kappa \left\langle \left|\delta
\theta_\kk\right|^2\right\rangle+\left\langle \frac{\delta
\theta_{-\kk} \delta n_\kk}{\bar n} \right\rangle \nonumber
\\ &+\frac{2(g-\kappa \tilde
\gamma)\bar n}{\epsilon_\kk}\left\langle \frac{\delta \theta_{-\kk}
\delta n_\kk}{\bar n} \right\rangle ,
 \label{dph2}
\end{align}
\begin{align}
\frac{D}{\bar n\epsilon_\kk}&= \left[\frac{\kappa}{2}+\frac{(\kappa
g+\tilde\gamma)\bar n}{\epsilon_\kk}\right]\left\langle
\left|\frac{\delta n_\kk}{\bar n}\right|^2\right\rangle \nonumber
\\ &-\left\langle
\frac{\delta \theta_{-\kk} \delta n_\kk}{\bar n} \right\rangle,
 \label{dn2}
\end{align}
\begin{align}
&\left[\epsilon_\kk +2(g-\kappa \tilde \gamma)\bar
n\right]\left\langle \left|\frac{\delta n_\kk}{\bar
n}\right|^2\right\rangle =4\epsilon_\kk\left\langle \left|\delta
\theta_\kk\right|^2\right\rangle  \nonumber
\\ &-4\left[\kappa\epsilon_\kk+
(\kappa g+\tilde\gamma)\bar n\right]\left\langle \frac{\delta
\theta_{-\kk} \delta n_\kk}{\bar n} \right\rangle,
 \label{dphn}
\end{align}
where
\begin{align}
\epsilon_\kk=k^2.
 \label{eps}
\end{align}

Eqs. \eqref{dph2}-\eqref{dphn} can be solved for the density and
phase fluctuations and are accurate when they are small. Close to
the BKT transition, this condition however breaks down. In the
following, we will outline how these equations can still be used in
order to obtain an estimate for the critical point, in analogy with
our study of the BKT transition in photon condensates
\cite{gladilin2021vortex}.

\section{Approximations for the BKT critical point \label{sec:approx}}

\subsection{Heuristic estimate of density-phase correlator \label{subsec:heur}}

In order to obtain our estimate of the critical point, we start by
integrating Eq. \eqref{dph2} over all momenta. In the right hand
side, we then use that for a homogeneous system
\begin{align}
    \int d^2\kk \langle  |\delta \theta_\kk|^2 \rangle &= \langle \delta \theta(\xx)\, \delta \theta (\xx) \rangle \equiv \langle \delta \theta^2 \rangle  \\
    \int d^2\kk \langle  \delta \theta_{-\kk} \delta n_\kk \rangle &= \langle \delta \theta(\xx)\, \delta n (\xx) \rangle \equiv \langle \delta \theta \delta n \rangle
\end{align}

When integrating the left-hand side of Eq.~(\ref{dph2}) over $\kk$,
we assume the presence of a finite UV momentum (energy) cutoff $k_+$
($\epsilon_+=k_+^2$). Our numerical simulations are performed for a
lattice with grid size $h$, for which our UV cutoff equals
$k_+=\pi/h$ [i.e, $\epsilon_+=(\pi/h)^2$]. Furthermore, one has to
take into account that for the systems, described by nonlinear
equations similar to Eq.~(\ref{ggpe2}), the use of the linear
approximation given by Eq.~(\ref{eps}) is physically
meaningful~\cite{prokof2001critical,gladilin2021vortex} only for $k$
above a certain IR momentum (energy) cutoff $k_-$
($\epsilon_-=k_-^2$). Then the Fourier transform of the left-hand
side of Eq.~(\ref{dph2}) can be represented as
$D[C_1+\ln(\epsilon_+/\epsilon_-)]/(4\pi\bar n)$, where the fitting
constant $C_1$ approximates the contribution of momenta smaller than
$k_-$.

{Physically, the correlator $\left\langle \delta \theta \delta n
\right\rangle$ expresses correlations between the density and
current fluctuations (since the velocity is the spatial derivative
of the phase). In nonequilibrium condensates, density and velocity
fluctuations are correlated because the particle balance equation: a
local suppression of the density leads to local reduction of
particle losses, which is compensated by an outward flow of
particles. In the context of the BKT transition, this physics plays
an important role, because the density in a vortex core is reduced
so that  vortices are accompanied by outgoing radial currents. The
magnitude of the density-phase correlator } was estimated in
Ref.~\cite{gladilin2021vortex} for nonequilibrium photon
condensates. Following this approach, for the system under
consideration here, we obtain
\begin{equation}
\langle \delta \theta \, \delta n \rangle = \frac{\tilde
\gamma}{\bar n} \langle \delta N^2 \rangle,
\end{equation}
where $\delta N = \int_0^x \delta n(x') dx'$. In the case of a plane
density wave $n=\bar n (1-a\cos kx )$ one has
\begin{equation}
\langle \delta N^2\rangle=\frac{a^2\bar n^2}{2k^2}. \label{dN2}
\end{equation}
At the BKT transition, vortices have to nucleate, which requires in
a continuum model strong density fluctuations with amplitude $\bar
n$ (i.e. $a = 1$) \cite{gladilin2021vortex}. Those strong
fluctuations have appreciable probability only for relatively large
momenta $k\sim k_+$ as seen from the fact that the best fitting in
Ref.~\cite{gladilin2021vortex} corresponds to the effective momentum
value $k\approx 0.3 k_+$ in Eq.~(\ref{dN2}). Therefore, we
approximate the correlator $\left\langle \delta \theta \delta n
\right\rangle$ by $C_2\bar n\tilde \gamma/\epsilon_+$, where
$C_2\sim 1$ is a fitting parameter.

Analogously, the Fourier transform of $\left\langle \delta
\theta_{-\kk} \delta n_\kk \right\rangle/\epsilon_\kk$ in the last
term of Eq.~(\ref{dph2}) is approximated by $C_3\bar n\tilde
\gamma/\epsilon_+^2$ with a fitting constant $C_3$. As a result, we
obtain the following approximate expression for the critical noise
\begin{align}
d_{\rm BKT}&=\left\{2\kappa \langle \delta \theta^2 \rangle_{\rm
BKT}+\left[C_2+\frac{2C_3(g-\kappa\tilde\gamma)}{\epsilon_+}\right]
\frac{\tilde\gamma}{\epsilon_+} \right\}\nonumber \\
&\times\frac{4\pi}{C_1+\ln(\epsilon_+/\epsilon_-)},
 \label{dBKTs}
\end{align}
where $d_{\rm BKT} \equiv \left.(D/\bar n)\right|_{\rm BKT}$.

In line with Refs.~\cite{prokof2001critical,gladilin2021vortex}, we
will assume that at the transition $\langle \delta \theta^2
\rangle_{\rm BKT}=1/2$. In the equilibrium case (and at $\kappa^2\ll
1$) the IR momentum cutoff is inversely proportional to the healing
length, so that the corresponding energy cutoff is $\sim g\bar n$.
Since the healing length corresponds at equilibrium to the vortex
core size, a natural generalization to the nonequilibrium situation
is to take a cutoff based on an estimate of the vortex core size.
Our estimation of the vortex core size, detailed in appendix
\ref{app:vort}, leads to
\begin{align}
\epsilon_-=\bar n \left[g+B_0\tilde \gamma \left(\frac{B_0\tilde
\gamma }{g+B_0\tilde \gamma}\right)^3 \right],
 \label{epslow1}
\end{align}
where $B_0=0.524$. The average density $\bar n$ in
Eq.~(\ref{epslow1}) will be approximated by its steady-state value
in the absence of noise \eqref{barn}.

\begin{figure} \centering
\includegraphics[width=1.0\linewidth]{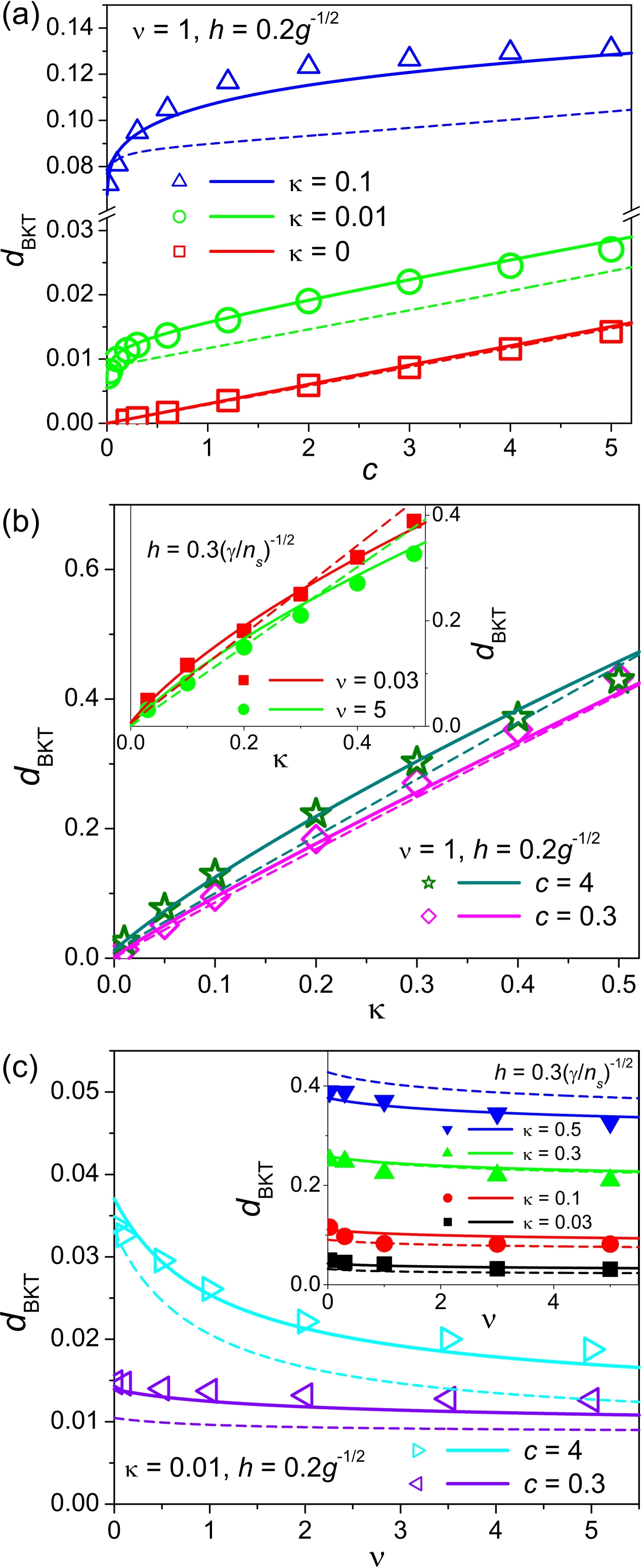}
\caption{Numerically (symbols) and semi-analyticaly (lines)
determined renormalized critical noise $d_{\rm BKT}=D_{\rm
BKT}/n_{\rm BKT}$ as a function of $c=\gamma/(2n_s g)$ (a), $\kappa$
(b), and $\nu$ (c). The insets in panels (b) and (c) show the
dependence of $d_{\rm BKT}$ on $\kappa$ and $\nu$, respectively, in
the case of $g=0$. The solid and dashed lines correspond to
Eqs.~(\ref{dBKT}) and (\ref{dBKTs}), respectively.
 \label{figfit}}
\end{figure}

\begin{figure} \centering
\includegraphics[width=1.\linewidth]{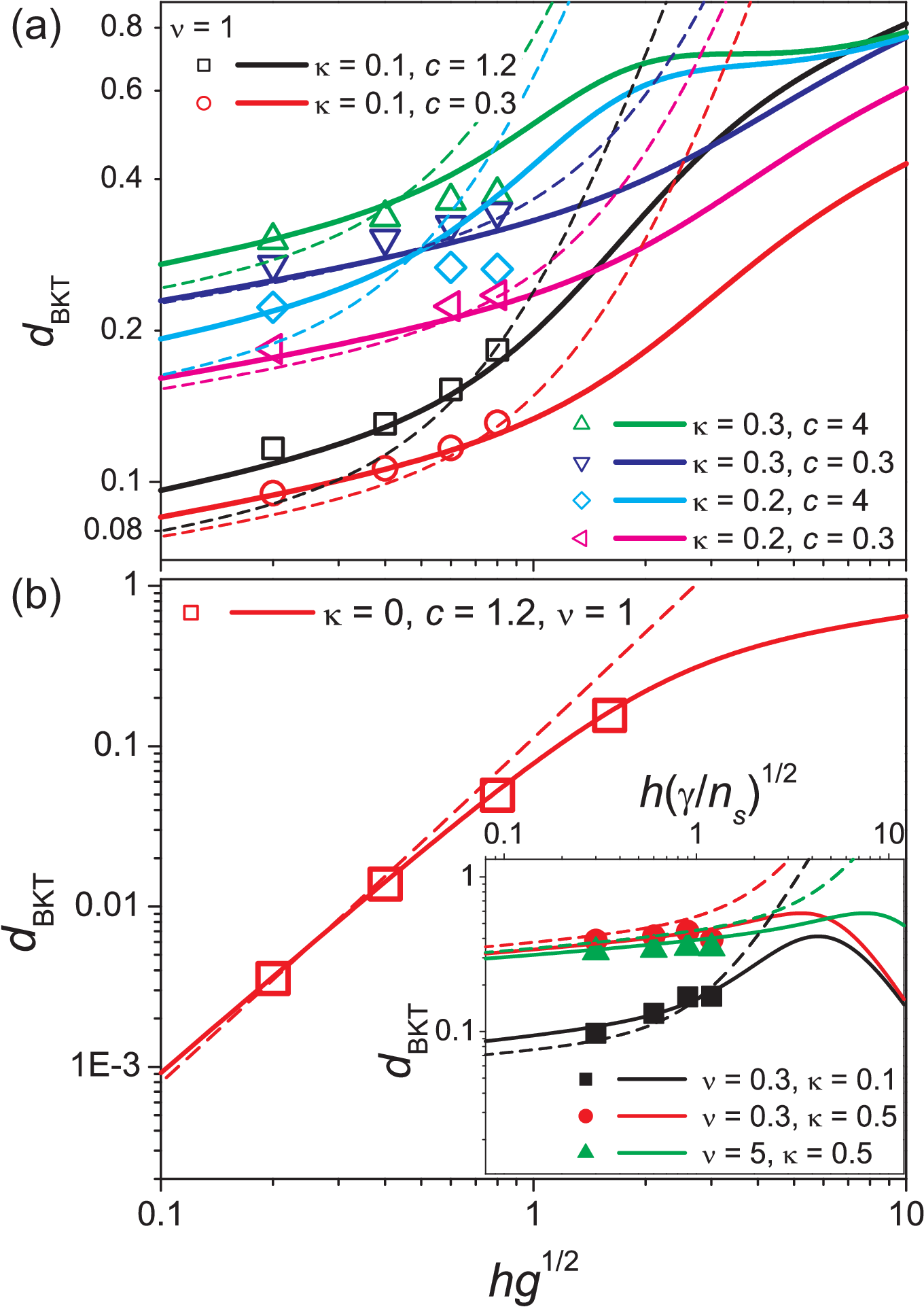}
\caption{Numerically (symbols) and semi-analyticaly (lines)
determined renormalized critical noise $d_{\rm BKT}$ as a function
of the grid step at $\kappa \geq 0.1$ (a) and $\kappa =0$ (b) for
nonzero $g$. Inset in panel (b): $d_{\rm BKT}$ as a function of the
grid step at $g=0$. The solid and dashed lines correspond to
Eqs.~(\ref{dBKT}) and (\ref{dBKTs}), respectively.
 \label{figh}}
\end{figure}

The results of fitting the numerical data for $d_{\rm BKT}$ with
Eq.~(\ref{dBKTs}) are represented by the dashed lines in
Figs.~\ref{figfit} and \ref{figh} where the determined fitting
parameters are $C_1=8.87$, $C_2=1.64$, and $C_3=5.92\times 10^{-5}$.
The small numerical value of $C_3$ implies it can actually be set to
zero without affecting the quality of the fits. {The numerical data
in Figs.~\ref{figfit}(a) and \ref{figh}(a) and the main panels in
Figs.~\ref{figfit}(b) and \ref{figh}(b) are taken from Ref.
\cite{gladilin2019noise-induced}. To numerically solve Eq.
(\ref{ggpe2}), a finite-difference scheme was used. Specifically, we
use periodic boundary conditions for a square of size $L_x=L_y=40$
with grid step equal to 0.2. The location of the critical point is
determined in the following way: after a long time evolution in the
presence of noise, the system was evolved without noise for a short
time (few our units of time) before checking for the presence of
vortices. This noiseless evolution gives the advantage of cleaning
up the density and phase fluctuations while it is too short for the
unbound vortex-antivortex pairs to recombine. The propensity for
their recombination is reduced \cite{gladilin2019noise-induced} with
respect to the equilibrium case thanks to outgoing radial currents
that provide an effective repulsion between vortices and
antivortices. To determine the critical noise for the BKT
transition, $D_{\rm BKT}$, we use the following criterion. If for a
noise intensity $D$ unbound vortex pairs are present after a noise
exposure time $t_D$ (and hence $D>D_{\rm BKT}$), while for a certain
noise intensity $D^\prime<D$ no vortex pairs appear even at noise
exposures few times longer then $t_D$, then $D^\prime$ lies either
below $D_{\rm BKT}$ or above $D_{\rm BKT}$ and closer to $D_{\rm
BKT}$ then to $D$. Therefore, the critical noise intensity can be
estimated as $D_{\rm BKT}=D^\prime \pm (D-D^\prime)$. }

As seen from the comparison between the dashed lines and the symbols
in Figs.~\ref{figfit} and \ref{figh}, Eq.~(\ref{dBKTs})
qualitatively reproduces the main trends in the behavior of the
numerically determined $d_{\rm BKT}(c,\kappa,\nu,h)$ at relatively
small grid steps $h$, when $\epsilon_+$ is considerably larger than
$\epsilon_-$. {This qualitative agreement is ensured, in particular,
by taking into account the contributions related to density-phase
correlation, which are zero in equilibrium systems but play a
crucial role for the BKT transition out of equilibrium. At the same
time, this simple and transparent heuristic estimate of these
contributions does not appear sufficient for a good quantitative
description of the numerical results.}

\subsection{Bogoliubov theory with nonlinear correction}

In order to obtain a better quantitative description of the numerics
for the nonequilibrium BKT transition, we develop below a different
approach that leads to a slightly more involved expression. To this
purpose, we start from the linear approximation for the phase
fluctuations in the steady state, obtained by solving Eqs.
\eqref{dph2}-\eqref{dphn}. Inserting ${D}/{\bar n}$ from
Eq.~(\ref{dph2}) and $\left\langle \left|{\delta n_\kk}/{\bar
n}\right|^2\right\rangle$ from Eq.~(\ref{dphn}) into
Eq.~(\ref{dn2}), we obtain the relation
\begin{align}
&\left[\epsilon_\kk+3g \bar n +2\left(g^2+\tilde\gamma^2\right)
\frac{\bar n^2}{\epsilon_\kk}\right]\left\langle \frac{\delta
\theta_{-\kk} \delta n_\kk}{\bar n} \right\rangle \nonumber
\\ & =2\tilde\gamma\bar n\left\langle
\left|\delta \theta_\kk\right|^2\right\rangle .
 \label{ph2phn}
\end{align}
Using Eq.~(\ref{ph2phn}), we express $\left\langle {\delta
\theta_{-\kk}\delta n_\kk}/{\bar n} \right\rangle $ through $\large
\langle \left|\delta \theta_\kk\right|^2\large\rangle $ and insert
the result into Eq.~(\ref{dph2}). For the phase fluctuations, this
leads to the equation
\begin{align}
&\left\langle \left|\delta
\theta_\kk\right|^2\right\rangle=\frac{D}{\bar n}f(\epsilon_\kk),
 \label{ph2D}
\end{align}
where
\begin{align}
 f(\epsilon)=\frac{1}{2\kappa}\frac{\epsilon+3\bar n g
 +2\left(g^2+\tilde\gamma^2\right)
 \bar n^2/\epsilon}{(\epsilon+\epsilon_1)(\epsilon+\epsilon_2)}.
 \label{f}
\end{align}
with
\begin{align}
\epsilon_1= \bar n\left(g+\frac{\tilde \gamma}{\kappa}\right),\quad
\epsilon_2= 2\bar n g.
 \label{eps12}
\end{align}
{From Eqs. \eqref{ph2D} and \eqref{f}, one sees that the phase
fluctuations are, as expected, proportional to the noise strength
$D$ and decrease as a function of the density $\bar n$ and energy
relaxation $\kappa$. For what concerns their energy dependence, Eq.
\eqref{f} shows a $1/\epsilon$ behavior both at small and large
energies. As a consequence, the Fourier transform of phase
fluctuations, needed to obtain their real space correlations
requires the introduction of an infrared cutoff $\epsilon_-$,
analogous to the treatment in Sec. \ref{subsec:heur}.  } As a result
of Fourier transformation, the local phase variance becomes
\begin{align}
\left\langle \delta \theta^2\right\rangle=\frac{D}{4\pi\bar
n}(F+F_-)
 \label{ph2Dr}
\end{align}
where
\begin{align}
F=&\int \limits_{\epsilon_-}^{\epsilon_+}f(\epsilon)d\epsilon
=\frac{1}{2}\frac{g^2+\tilde \gamma^2}{g(\kappa g+\tilde
\gamma)}\ln\left(\frac{\epsilon_+}{\epsilon_-}\right) \nonumber
\\&+\frac{\tilde\gamma}{\tilde\gamma+\kappa g}\left(\frac{1}{2\kappa}
+\frac{\kappa\tilde \gamma}{\tilde \gamma-\kappa
g}\right)\ln\left(\frac{\epsilon_+ +\epsilon_1}{\epsilon_-
+\epsilon_1}\right) \nonumber
\\&-\frac{\tilde \gamma^2}{2g(\tilde \gamma-\kappa
g)}\ln\left(\frac{\epsilon_+ +\epsilon_2}{\epsilon_-
+\epsilon_2}\right),
 \label{Flin}
\end{align}
{where the logarithmic dependence on the lower and upper energy
cutoffs is a consequence of the $1/\epsilon$ behavior of
$f(\epsilon)$ at low and high energies. } The term
\begin{align}
F_-=C_-\epsilon_- f(\epsilon_-)
 \label{Fminus}
\end{align}
in Eq.~(\ref{ph2Dr}) approximates the contribution of the integral
over $\epsilon$ from 0 to $\epsilon_-$, where $C_-$ is a fitting
parameter.

Expression (\ref{ph2Dr}), derived with the use of linearized
equations for the phase and density fluctuations, is expected to be
applicable when these fluctuations are small. As discussed above, at
the BKT transition, where both phase and density fluctuations are
large, the real-space correlator $\left\langle \delta \theta \delta
n \right\rangle$ is mainly determined by the contributions of $k\sim
k_+$. According to Eq.~(\ref{ph2phn}), the quantity $\large\langle
\left|\delta \theta_\kk\right|^2\large\rangle$ contains a term that
is exactly proportional to $\left\langle {\delta \theta_{-\kk}
\delta n_\kk} \right\rangle$. This implies that at the BKT
transition the expression for the phase fluctuations $\large\langle
\delta \theta^2\large\rangle$, derived above, needs an additional
``nonlinear correction'', which would describe an enhanced
contribution of large momenta $k\sim k_+$ (large energies
$\epsilon\sim \epsilon_+$). Here, we approximate this correction by
adding to $F$ the term
\begin{align}
F_+&= C_+ \epsilon_+\, f(\epsilon_+),
 \label{Fplus}
\end{align}
where $C_+$ is a fitting parameter. Then at the BKT point we have
\begin{align}
d_{\rm BKT}= \langle \delta \theta ^2\rangle_{\rm BKT} \; \frac{4\pi
}{F+F_-+F_+},
 \label{dBKT}
\end{align}
{where again we take $\langle \delta \theta ^2\rangle_{\rm
BKT}=1/2$.}

Applying Eq.~(\ref{dBKT}) to fit the numerical data for $d_{\rm
BKT}$, we obtain for the two fitting parameters: $C_-=2.24$ and
$C_+=7.33$. {As compared to the results of the heuristic approach
described in the previous subsection (dashed lines in
Figs.~\ref{figfit} and \ref{figh}), the results corresponding to
more involved and accurate Eq.~(\ref{dBKT}), which are shown by the
solid lines in Figs. 1 and, demonstrate a much better quantitative
agreement with the numerically determined $d_{\rm BKT}$. }

\begin{figure} \centering
\includegraphics[width=1.0\linewidth]{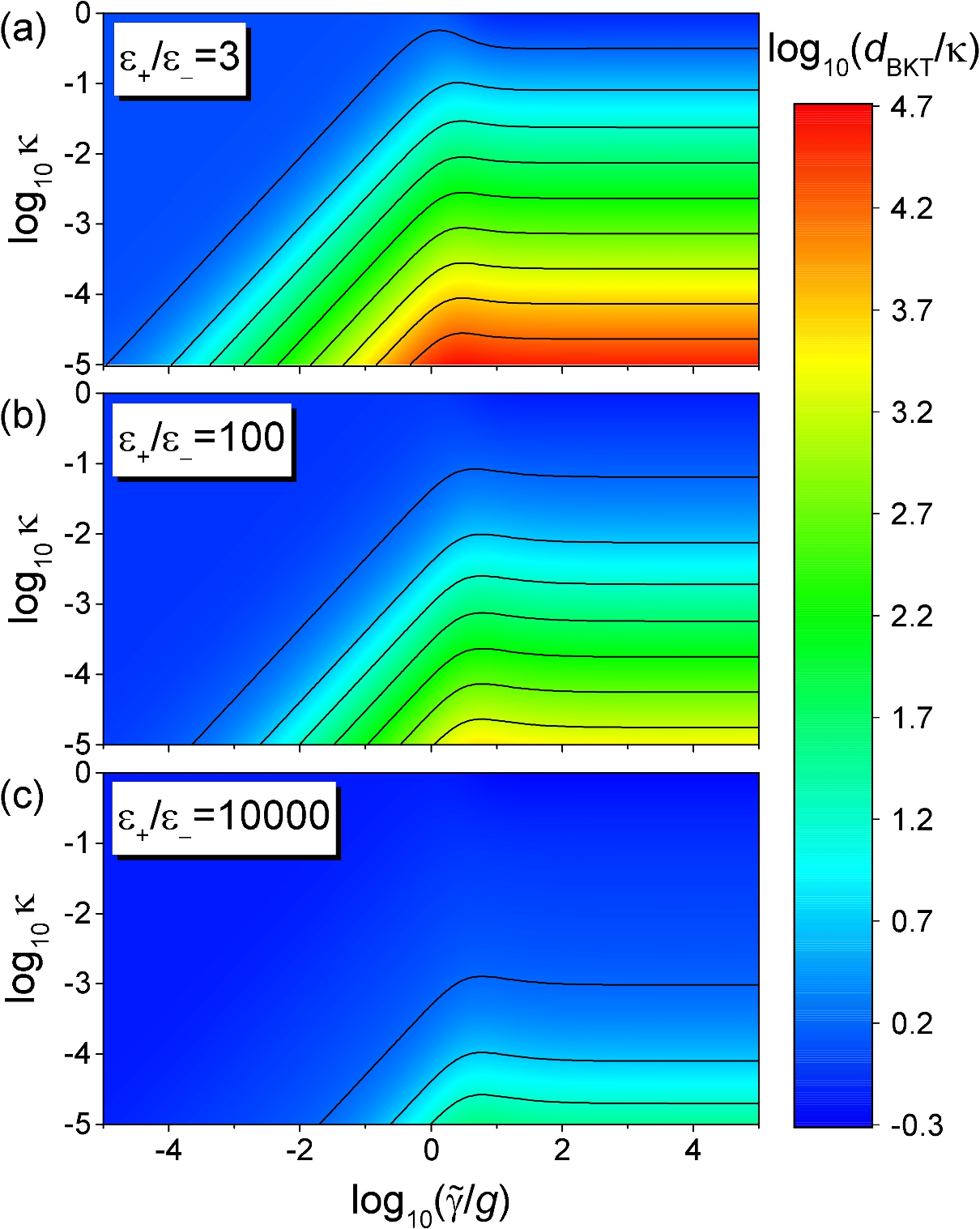}
\caption{Renormalized critical noise $d_{\rm BKT}/\kappa$, given by
Eq.~(\ref{dBKT}), as a function of $\tilde\gamma/g$ and $\kappa$ at
three different values of $\epsilon_+/\epsilon_-$.
 \label{dmap}}
\end{figure}

The semi-analytical expression for $d_{\rm BKT}$, given by
Eq.~(\ref{dBKT}) together with Eqs. (\ref{epslow1}), (\ref{f}),
(\ref{eps12}), and (\ref{Flin})-(\ref{Fplus}), can be considered as
a function of three independent parameters: $\tilde\gamma/g$,
$\kappa$ and $\epsilon_+/\epsilon_-$. In Fig.~\ref{dmap}, the
renormalized critical noise $d_{\rm BKT}/\kappa$, corresponding to
Eq.~(\ref{dBKT}), is plotted for a wide range of the parameters
$\tilde\gamma/g$ and $\kappa$ at three different values of the ratio
$\epsilon_+/\epsilon_-$.

For small losses and not too small $\kappa$, the ratio $d_{\rm
BKT}/\kappa$ is of order one, in line with the equilibrium BKT
transition where according to fluctuation-dissipation relation $D =
\kappa T$ \cite{hohenberg1977theory} and where the critical
temperature scales in first approximation as $T_{BKT} \sim n$. In
line with our previous studies for polariton condensates
\cite{gladilin2019noise-induced} and photon condensates
\cite{gladilin2021vortex}, we see that the losses stabilize the
ordered phase: when $\tilde \gamma$ is increased at fixed $\kappa$,
the noise required to make the transition to the state with free
vortex-antivortex pairs increases. We explained this trend by the
reduction of the density fluctuations for increased driving and
dissipation \cite{gladilin2019noise-induced}, that manifests itself
through density-phase correlations \cite{gladilin2021vortex} [see
discussions preceding Eq. \eqref{dBKTs} and Eq. \eqref{Fplus}].

In the limit without losses ($\tilde \gamma = 0$), our estimate for
the critical point reduces to
\begin{equation}
n_{\rm BKT} =  \frac{T_{\rm BKT}}{2\pi} \left[ \log \left(
\frac{1}{m h^2 g n_{\rm BKT}} \right) + A_1 \right].
\end{equation}
Here, we have used that $T_{\rm BKT} = D_{\rm BKT}/\kappa$, defined
$A_1 =  C_+ + C_- + \log(\pi^2/2) \approx 11.2$ and restored
physical units. We can compare this expression with the equilibrium
BKT transition for the weakly interacting lattice Bose gas (Eq. (12)
in \cite{prokof2001critical})
\begin{equation}
    n_{\rm BKT} = \frac{m T_{\rm BKT}}{2 \pi} \log \frac{A}{mh^2 g T_{\rm BKT} },
\end{equation}
with $A = 6080$. This expression can be written as
\begin{equation}
    n_{\rm BKT} = \frac{m T_{\rm BKT}}{2\pi} \left[ \log\left( \frac{1}{mh^2 g n_{\rm BKT}}\right) + A_2 \right],
\end{equation}
with
\begin{equation}
    A_2 =\log\left[ \frac{A}{2\pi} \log\left(\frac{A}{m^2 h^2 g T_{\rm BKT}}\right)  \right].
\end{equation}
Assuming here $m^2 h^2 g T_{\rm BKT} \approx 1$, one obtains
$A_2\approx 9.1$, which is reasonably close to our $A_1  \approx
11.5$ given the simplicity of our approach and considering that the
equilibrium case is actually a somewhat singular limiting case of
our model where the gain and losses simultaneously tend to zero.

\section{Conclusions \label{sec:concl}}

In this paper, we have developed a semi-analytical approach to
describe the BKT transition point for driven-dissipative weakly
interacting Bose gases. We start from the linearized equations of
motion for the density and phase fluctuations and subsequently
correct phenomenlogically for nonlinearities that are important
close to the BKT transition. Our resulting analytical formulae
contain some fitting parameters that are fitted to a series of
numerical simulations in a wide parameter range. The good fitting of
our numerical results indicates the validity of the physical
intuition underlying our semi-analytical approach and promotes our
formulae to a concise summary of the numerical results.

Of course, our numerical results were obtained for a finite size
system and we can therefore not settle what will happen for much
larger system sizes, where it remains possible that the KPZ
nonlinearity may destabilize the algebraically ordered phase
\cite{altman2015two,wachtel2016electrodynamic}, even though recent
numerical work has shown that KPZ scaling can be witnessed in 2D
nonequilibrium condensates without the phase coherence being
destabilized by the formation of vortex antivortex pairs
\cite{mei2021spatiotemporal,deligiannis2022kardar}.

\section*{Acknowledgements}
We thank Iacopo Carusotto for continuous stimulating discussions. VG
was financially supported by the FWO-Vlaanderen through grant nr.
G061820N.

\input{output.bbl}

\appendix
\section{Vortex density profile \label{app:vort}}

The vortex core size plays an important role in the BKT physics,
because it provides the low energy cutoff in our analytical
treatment. In this appendix, we discuss how the vortex core size
depends on the system parameters through an approximate solution of
the gGPE, that is shown to compare favorably with the exact
numerical solution.

We consider a single-quantum vortex in an infinite 2D condensate.
Assuming that the vortex-center position is fixed, the density
distribution is circularly symmetric and the order parameter can be
written in the cylindrical coordinates $\rho$ and $\phi$ as
$\psi=\chi(\rho)e^{-i\phi}$, so that the condensate density is given
by $n=|\chi|^2$. Inserting this into the noise-free form of
Eq.~(\ref{ggpe2}), one has
\begin{align}
\frac{\partial\chi}{\partial t}=&(i+\kappa)
\left[\frac{\partial^2}{\partial\rho^2}+\frac{1}{\rho}
\frac{\partial}{\partial\rho}-\frac{1}{\rho^2} -g|\chi|^2\right.
\nonumber
\\
&\left.+\frac{i\gamma}{2n_s}\frac{1-|\chi|^2}{1+\nu |\chi|^2}
\right] \chi . \label{Achi}
\end{align}
For analytical estimates it is convenient to represent $\chi$ as
$\chi(\rho)=\sqrt{\bar n}y(\rho)e^{i\theta(\rho)}$, where the real
function  $y(\rho)$ is normalized by 1. Then, taking into account
that for a steady state ${\partial y}/{\partial t}=0$, while
${\partial \theta}/{\partial t}= -\mu (1+\kappa^2)$ with $\mu$, the
chemical potential, one obtains from Eq. (\ref{Achi}) the following
two coupled stationary differential equations:
\begin{align}
\kappa \mu=&\frac{\gamma}{2n_s}\frac{1-\bar n y^2}{1+\nu \bar n y^2}
-\frac{1}{\rho y^2}\frac{\partial}{\partial \rho}\left(\rho y^2
\frac{\partial \theta}{\partial \rho}\right),
 \label{Ae1}
\end{align}
\begin{align}
\frac{1}{\rho^2}-\frac{1}{\rho y}\left(\rho \frac{\partial
y}{\partial \rho}\right)=&\mu -\left(\frac{\partial \theta}{\partial
\rho}\right)^2-g\bar n y^2.
 \label{Ae2}
\end{align}
In Eq.~(\ref{Ae2}), the first term  corresponds to circulating
vortex flows, while the second term in the right-hand side is due to
outward radial flows from the vortex core
\cite{gladilin2017interaction}.

Considering Eq.~(\ref{Ae2}) in the limit $\rho\to \infty$, one
obtains for the chemical potential
\begin{align}
\mu= \left.\left(\frac{\partial \theta}{\partial
\rho}\right)^2\right|_{\rho \to \infty} +g\bar n.
 \label{Amu}
\end{align}
Note that in the equilibrium case, when ${\partial \theta}/{\partial
\rho}=0$, the right hand side of Eq.~(\ref{Ae2}) is obviously
positive. In order to keep it positive also far from equilibrium,
one has to assume that $\left.\left({\partial \theta}/{\partial
\rho}\right)^2\right|_{\rho \to \infty}$ is nonzero. In other words,
in the presence of a vortex the chemical potential of a
nonequilibrium system should increase.

In the limit $\rho\to 0$, when $({\partial \theta}/{\partial
\rho})^2$ and $y^2$ become negligibly small, the general
non-divergent solution of the ``reduced'' equation, resulting from
Eq.~(\ref{Ae2}), is simply $CJ_1(q\rho)$, where $J_1(x)$ is the
Bessel function and $q=\sqrt{\mu}$. Let us consider the
``equilibrium-like'' version of Eq.~(\ref{Ae2}):
\begin{align}
\frac{1}{\rho^2}-\frac{1}{\rho y}\left(\rho \frac{\partial
y}{\partial \rho}\right)=&\mu(1- y^2).
 \label{Ae2b}
\end{align}
Its solution can be approximated by the normalized by one,
non-oscillating function
\begin{align}
y_1(\rho)=\frac{1}{J_1(x_*)}J_1\left(\frac{x}{\sqrt{1+(x/x_*)^2}}\right),
 \label{Ay1}
\end{align}
where $x=sq\rho$. The parameters $s$ and $x_*$ are determined from
the following two requirements. (i) At small $\rho$, the function
$y_1(\rho)$ should coincide with $CJ_1(q\rho)\approx
C[q\rho/2-(q\rho)^3/16]$. This leads to $s=(1+4/x_*^2)^{-1/2}$. (ii)
$y_1(\rho)$ should satisfy Eq.~(\ref{Ae2b}) in the limit $\rho\to
\infty$. In this limit, one has $1-y_1(\rho)\propto \rho^{-2}$ and
Eq.~(\ref{Ae2b}) becomes
\begin{align}
\frac{1}{\rho^2}=&\mu \frac{x_*^3 J_1^\prime (x_*)}{(sq\rho)^2
J_1(x_*)},
 \label{Ae2c}
\end{align}
leading for $x_*$ to the equation ${J_1^\prime
(x_*)}\left(x_*^3+4x_*\right)={J_1(x_*)}$, which gives $x_*=1.72$
and, correspondingly, $s=0.653$. As we will see later, in the case
of weak non-equilibrium, the function
\begin{align}
n_1(\rho)=\bar n y_1^2(\rho)
 \label{An1}
\end{align}
describes almost perfectly the vortex density profiles, found in
numerical simulations. Moreover, close to the vortex center, this
function works quite well even at relatively strong deviations from
equilibrium. This is not surprising: close to the vortex center, the
vortex circulating-current density, which is proportional to
$1/\rho$, is much stronger than the radial-current density, so that
just the former governs the particle-density suppression.

Let us estimate ${\partial \theta}/{\partial \rho}$, which
determines the radial particle flow. At $\rho \to \infty$, the last
term of Eq. (\ref{Ae1}) (which is proportional to ${\rm div}
j_\rho$) vanishes, while $y$ goes to 1, so that we have
\begin{align}
\kappa \mu=&\frac{\gamma}{2n_s}\frac{1-\bar n}{1+\nu \bar n}.
 \label{Ae1a}
\end{align}
Therefore, Eq. (\ref{Ae1}) can be rewritten as
\begin{align}
\frac{1}{\rho y^2}\frac{\partial}{\partial \rho}\left(\rho y^2
\frac{\partial \theta}{\partial \rho}\right)=&\tilde \gamma \bar
n\frac{1-y^2}{1-p(1- y^2)}
 \label{Adivj}
\end{align}
with $p={\nu \bar n}/(1+\nu\bar n)$. From Eq. (\ref{Adivj}) one
obtains
\begin{align}
\frac{\partial \theta}{\partial \rho}=& \frac{\tilde \gamma \bar
n}{sq} Q_p(\rho),
 \label{Adthdr}
\end{align}
where
\begin{align}
Q_p(\rho)=& \frac{sq}{\rho y^2(\rho)} \int \limits_0^\rho
d\rho^\prime \rho^\prime
\frac{y^2(\rho^\prime)\left[1-y^2(\rho^\prime)\right]}{1-p[1-
y^2(\rho^\prime)]}.
 \label{AQp}
\end{align}
 A finite nonzero value of $\left.{\partial \theta}/{\partial
\rho}\right|_{\rho \to \infty}$ is possible only if we assume that
at $\rho \to \infty$
\begin{align}
n(\rho)=\bar n y^2(\rho)\approx \bar n
\left(1-\frac{R}{\rho}\right).
 \label{Ay2inf}
\end{align}
Then we have from Eqs. (\ref{Adthdr}) and (\ref{AQp})
\begin{align}
\left.\frac{\partial \theta}{\partial \rho}\right|_{\rho \to
\infty}=R \tilde \gamma \bar n.
 \label{Adthdr2}
\end{align}
At moderate distances from the vortex center, the radial current
density increases with $\rho$. For sufficiently large $\tilde
\gamma$, the suppressive effect of redial currents on $y^2$ becomes
dominating above certain $\rho$, so that the behavior described by
Eq. (\ref{Ay2inf}) emerges.

\begin{figure} \centering
\includegraphics[width=1.0\linewidth]{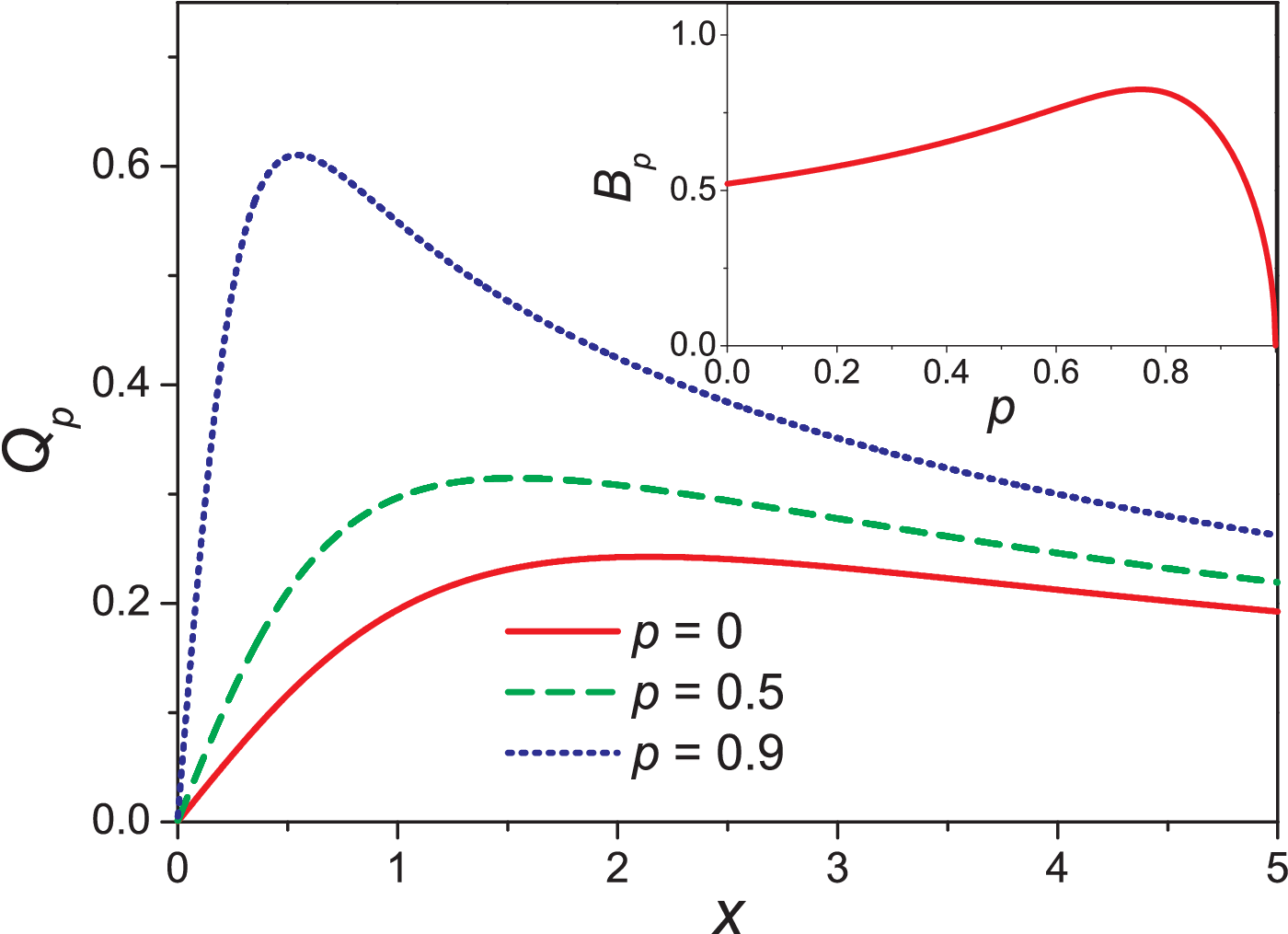}
\caption{Function $Q_p(x)$ with $y=y_1$ for three different values
of $p$. Inset: parameter $B_p$ as a function of $p$.
 \label{fQp}}
\end{figure}

In order ro determine the parameter $R$, let us consider the
crossover between the two regimes, described by Eqs. (\ref{An1}) and
(\ref{Ay2inf}). Let us start with the case of noninteracting
particles, $g=0$. The suppressive effect of the radial currents on
the particle density is determined by $({\partial \theta}/{\partial
\rho})^2$. At $\rho$ below the crossover point, $y$ in Eq.
(\ref{AQp}) can be approximated by $y_1$, so that $Q_p$ depends on
$\rho$ only through $x$ (see Fig. \ref{fQp}).
It seems natural to expect that the crossover occurs at a distance
$\rho_c$, where the value of $Q_p(x)$ is close to its maximum. For
simplicity, we will assume that the crossover point $\rho_c(p)$ just
corresponds to the position of this maximum, $x_m(p)$, i.e
$\rho_c=x_m/(sq)$. At the crossover point, the solution $y_1$ for
small $\rho$ should match the solution for large $\rho$, described
by Eq. (\ref{Ay2inf}). This leads to
\begin{align}
R= \sqrt{\frac{B_p}{\tilde \gamma\bar n}}
 \label{AR}
\end{align}
with
\begin{align}
B_p= \frac{x_m}{s}\left[1-y_1^2(\rho_c)\right],
 \label{AB}
\end{align}
where, as seen from Eq.~(\ref{Ay1}), $y_1(\rho_c)$ is determined
solely by $x_m(p)$. The numerically determined dependence of $B_p$
on $p$ is shown in the inset of Fig.~\ref{AQp}.

We can expect that in the general case, where the interparticle
interaction is non-negligible, the crossover occurs when, with
increasing $\rho$, the density of the radial current becomes
comparable with that of the circulating current, so that [see
Eq.~(\ref{Adthdr})],
\begin{align}
\left(\frac{\tilde \gamma \bar n}{sq}\right) Q_p(\rho_c)=&
\frac{C}{\rho_c}.
 \label{Across2}
\end{align}
Obviously, with increasing $g$ the suppressive effect of radial
currents on the particle density becomes relatively weaker.
Therefore, $R$ should decrease with increasing $g$ or decreasing
$\gamma$ ($R=0$ at $\gamma=0$). This means that at non-negligible
$g$ the matching condition at the crossover point,
${R}/{\rho_c}=1-y_1^2(x_c)$, corresponds to a rather small value of
$1-y_1^2(x_c)$, which can be approximated [see Eqs. Eq.~(\ref{Ay1}),
(\ref{Ae2c})] by $1/(q\rho_c)^2$. Then the matching condition
becomes $1/\rho_c=Rq^2$. Inserting this into Eq. (\ref{Across2}),
we obtain
\begin{align}
R &=\frac{Q_p(\rho_c)}{sC}\frac{(\tilde \gamma \bar n)}{q^3}.
 \label{AR2}
\end{align}
For simplicity, in the denominator $q^3$ we approximate $R$ by the
value given by Eq. (\ref{AR}). The constant $C$ is determined by
requiring that in the limit $g\to 0$ the $R$, given by Eq.
(\ref{AR2}), fits Eq. (\ref{AR}). Then for $R$ we finally have
\begin{align}
R &=\sqrt{\frac{B_p}{\tilde\gamma\bar n}} \left(\frac{B_p\tilde
\gamma }{g+B_p\tilde \gamma}\right)^{3/2}.
 \label{ARg}
\end{align}

From Eqs. (\ref{Amu}) and (\ref{Adthdr2}) with (\ref{ARg}), we
 obtain the relation
\begin{align}
\mu=\bar n \left[g+B_p\tilde \gamma \left(\frac{B_p\tilde \gamma
}{g+B_p\tilde \gamma}\right)^3 \right].
 \label{Amufin}
\end{align}
Equations (\ref{Amufin}) and (\ref{Ae1a}) completely define the
chemical potential $\mu$ and average density $\bar n$, which,
together with the parameter $R$ given by Eq. (\ref{ARg}), enter the
density distributions (\ref{An1}) and (\ref{Ay2inf}) at small and
large $\rho$, respectively. As a ``smooth interpolation'' between
these distributions, we introduce the function
\begin{align}
n_2(\rho)=&\frac{1}{1+{R}/{\rho}}n_1(\rho).
 \label{n2}
\end{align}
\begin{figure} \centering
\includegraphics[width=1.0\linewidth]{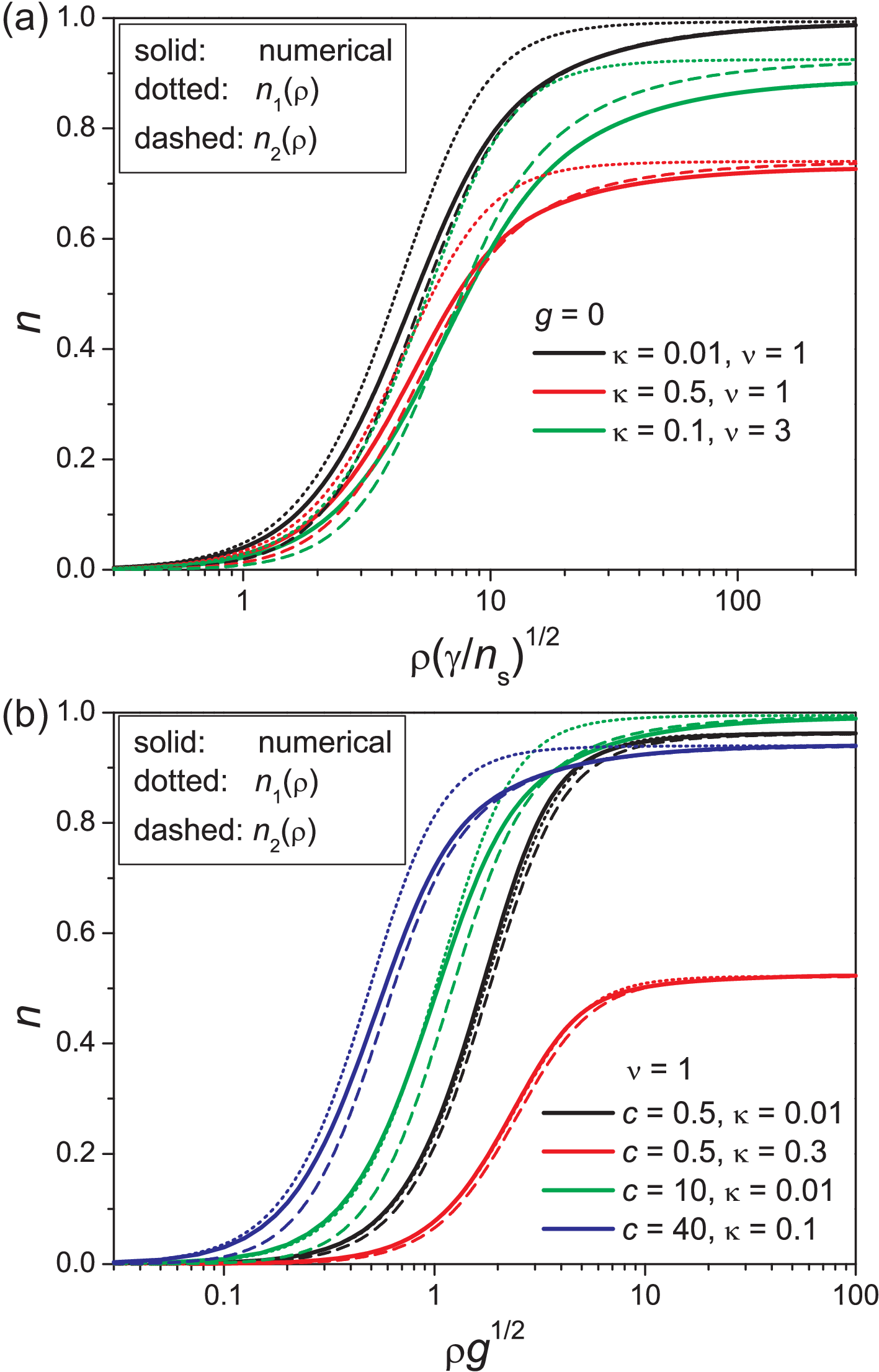}
\caption{Numerically (solid lines) and analytically (dotted and
dashed lines) calculated density profiles for noninteracting
particles (a) and three finite values of the parameter $c=\gamma/(2
n_s g)$ (b) at different $\nu$ and $\kappa$.
 \label{AFn}}
\end{figure}

Obviously, this function can somewhat underestimate $n$ at $\rho\sim
R$, close to the ``bottom'' of the vortex core. Apart from this, as
seen from Fig.~\ref{AFn}a, at $g=0$ the function $n_2(\rho)$
approximates rather well the vortex shape, found by solving
Eq.~(\ref{Ay1}) numerically, although the analytical values of $\bar
n$ appears not quite accurate for (experimentally less relevant)
large $\kappa$ (red curves) and large $\nu$ (green curves). For
strongly interacting particles and/or for week deviations from
equilibrium, when the parameter $c=\gamma/(2 n_s g)$ is smaller than
1, the numerical results are almost perfectly described by the
``equilibrium-like profile'' $n_1(\rho)$ (see the black and red
curves in Fig.~\ref{AFn}b). For $c>1$, the numerically determined
$n(\rho)$ at large $\rho$ is well approximated by $n_2(\rho)$ (see
the green and blue curves in Fig.~\ref{AFn}b).

The obtained results show that the $\mu$ given by Eq. (\ref{Amufin})
($q^{-1}=1/\sqrt{\mu}$) adequately describes the chemical potential
(vortex core size) in the systems under consideration. This implies
that Eq. (\ref{Amufin}) can provide a suitable estimate for the
lower energy cutoff $\epsilon_-$. Since for experimentally relevant
$p<0.9$ the parameter $B_p$ relatively weakly depends on $p$, in
this estimate, for simplicity, we replace $B_p$ with $B_0=0.524$.

\end{document}

%% file: output.bbl
%